\documentclass[%
 reprint,
 amsmath,amssymb,
 aps,
]{revtex4-2}

\usepackage{graphicx}
\usepackage{dcolumn}
\usepackage{bm}
\usepackage{hyperref}
\usepackage[mathlines]{lineno}

\usepackage{ amsmath,amssymb,amsthm,xcolor}
\usepackage{soul} 
\usepackage{tikz} 
\usepackage{orcidlink}

\newcommand{\ket}[1]{\left|#1\right>}

\newcommand{\expt}[1]{\left\langle #1 \right\rangle}
\newcommand{\Tr}{\text{Tr}}
\usepackage{xparse} 

\newcounter{example}[section]

\newcommand{\LE}{\text{LE}}

\newcommand{\avgconr}{\left<C(R_1,R_2)\right>}
\newcommand{\conr}{C(R_1,R_2)}

\newcommand{\orderp}{R}
\newcommand{\avgorderp}{\left\langle \orderp \right\rangle}
\newcommand{\corle}{C_{\text{LE}}}
\newcommand{\leref}{\text{LE}_{\text{ref}}}

\begin{document}

\preprint{APS/123-QED}

\title{Localizable Entanglement as an Order Parameter for Measurement-Induced Phase Transitions}

\author{Sourav Manna\orcidlink{0009-0005-5216-567X}}
\email{mannaphy@physics.iitm.ac.in}

 \author{Arul Lakshminarayan\orcidlink{https://orcid.org/0000-0002-5891-6017}}
 \email{arul@physics.iitm.ac.in}
 
 \author{Vaibhav Madhok\orcidlink{https://orcid.org/0000-0003-0785-0094}}%
 \affiliation{
 Department of Physics and Center for Quantum Information, Communication and Computing,
	Indian Institute of Technology Madras, Chennai, India 600036
}
 

\date{\today}

\begin{abstract}
We identify localizable entanglement (LE) as an order parameter for measurement-induced phase transitions (MIPT). LE exhibits universal finite-size scaling with critical exponents that match previous MIPT results and gives a nice operational interpretation connecting MIPTs to classical percolation.  Remarkably, we find that LE decays exponentially with distance in the area-law phase as opposed to being essentially constant for the volume-law phase thereby, discover an intrinsic length scale $\xi_E$ that diverges at the critical measurement probability $p_c$. While classical percolation transition captures successful transport across a network, MIPT as characterized by LE can be interpreted as quantifying the amount of quantum teleportation between two given nodes in a quantum circuit. Building on this insight, we propose a two-ancilla protocol that provides an experimentally accessible readout of entanglement redistribution across the transition.

\end{abstract}

\maketitle

The interplay between unitary dynamics and projective measurements gives rise to a new class of non-equilibrium phenomena known as measurement-induced phase transitions (MIPT) \cite{Skinner_2019, Fisher_2023, Iadecola_2023, Li_2019, PhysRevB.102.224311, PhysRevLett.132.110403, NahumRoyPRX2021, BuchholdAltladPrx2021, chaki2025MIPT,MannaPra2024,PiotrPrb2022,PiotrPrl2022}. In these hybrid quantum circuits, the rate of measurement controls a transition between a volume-law entangled phase and an area-law disentangled phase. This transition has been interpreted as a competition between entanglement generation by unitary evolution and information loss through local measurements, and has been studied in various random circuit architectures.

Despite extensive numerical and theoretical work, a central conceptual issue being vigorously pursued is the notion of correlation length and it's operational meaning in these measurement-driven systems. Progress has been made drawing analogies to classical percolation, expressing the scaling as $(p-p_c)^{-\nu}$. However, it is desirable to identify a physical quantity having a direct operational or physical meaning in the context of quantum information flow. 

In this work, we address this question by identifying localizable entanglement (LE) \cite{CiracPrl2004,VerstraetePrl2004,PoppPra2005,AmicoRmp2008} as a physically inspired correlation length in monitored quantum circuits.
The LE between two sites quantifies the maximal entanglement that can be concentrated between them through optimal local measurements on the rest of the system. This makes it a natural diagnostic of how far quantum information can propagate before being irreversibly lost to measurements. 
LE informs us about the distance over which entanglement can be accessed via local measurements.
Therefore, in models like the Heisenberg or Ising chains, LE helps characterize quantum phase transitions as this length  diverges as one approaches the critical point  \cite{VerstraetePrl2004, CiracPrl2004}.
Operationally, LE is a resource for quantum teleportation as it quantifies the entanglement that can be localized between distant parties via local measurements \cite{BarasinSRski2018}.
 
Using random Clifford circuits, we exploit the stabilizer–graph-state correspondence to compute LE exactly across system sizes and measurement rates. Our results show that the LE falls off exponentially with distance in the area-law phase, but stays nearly flat in the volume-law phase thereby defining an intrinsic length scale $\xi_E$ that diverges at the critical measurement probability $p_c$ providing an operational interpretation of a length scale: $\xi_E$ measures the spatial extent over which entanglement can be concentrated by local operations. To the best of our knowledge, this is the first time such an exponential decay has been explicitly demonstrated. 
The MIPT transition can be probed by employing a reference qubit  initially maximally entangled with one site of the circuit as it undergoes hybrid measurement-unitary dynamics. A finite LE between the reference and the system captures the volume-law phase and vanishes in the area-law phase, with the crossover sharpening with system size. This picture naturally captures the intuitive notion that below the critical measurement rate, the reference “communicates” with the system, while above it, the reference becomes isolated due to measurement-induced disentanglement.

Further exploiting the structure of correlations and monogamy of entanglement, we introduce a second reference qubit, initially unentangled with the system and the first reference qubit. The two reference qubits are then entangled by a two-qubit random gate after the monitored evolution. The resulting concurrence between the two references reflects the connectivity of the system: when the system remains volume-law entangled, monogamy exponentially suppresses the entanglement between the two references, while in the area-law phase, the references become directly entangled. Remarkably, the concurrence between the two references alone—without access to the system—faithfully detects the MIPT. This provides an experimentally feasible method for probing the transition based solely on two-qubit measurements.

Together, these results provide (i) an operational definition of entanglement correlation length through LE, and (ii) a monogamy-based, low-overhead probe of measurement-induced phases. They bring together percolation-based geometric connectivity, information-theoretic, and experimental perspectives on MIPT, offering a unified framework to study entanglement dynamics in monitored quantum systems.

\begin{figure}
    \centering
    \includegraphics[width=\linewidth]{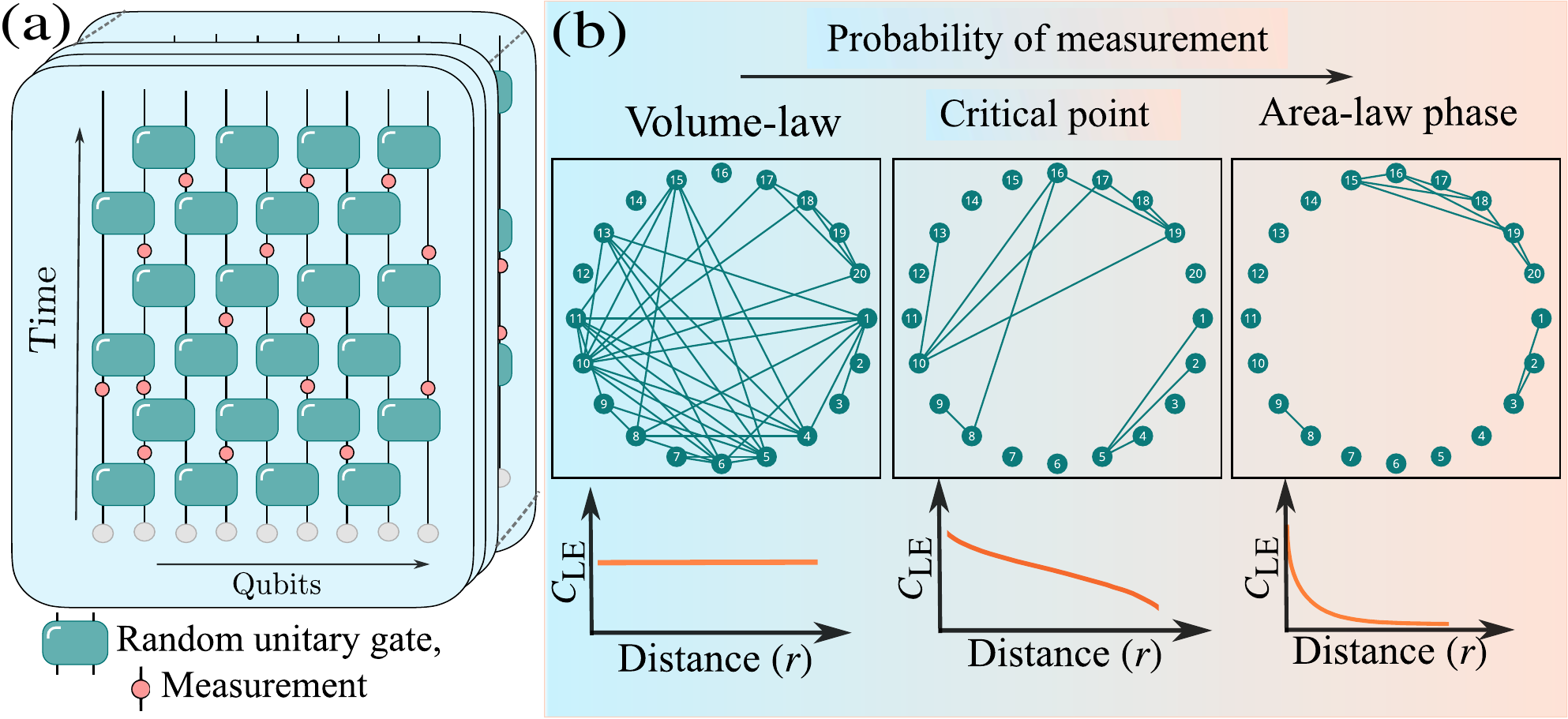}
    \caption{(a) Schematic of a $(1+1)$-dimensional monitored brickwall quantum circuit composed of discrete time layers of random unitary gates and projective measurements, exhibiting a measurement-induced transition. All quantities in the main text are obtained by averaging independent realizations over such circuits. (b) Graph state representations of stabilizer states generated from a 20-qubit MIPT circuit with Clifford gates evolved for a sufficiently long time at three representative measurement probabilities, arranged from left to right with increasing measurement rate: volume-law phase $(p = 0.14)$, critical point $(p=0.16)$, and area-law phase $(p=0.18)$. The graphs show the connectivity at each phase. The schematic shown below the graph state depicts the corresponding behavior of the averaged correlation function $\left<\corle \right>$: distance-independent in the volume-law phase, intermediate at criticality, and exponentially decaying in the area-law phase.}
    \label{fig:graphabs}
\end{figure}

The LE quantifies the maximum entanglement that can be concentrated between two selected qubits ($i$ and $j$) on average by performing local measurements on the remaining qubits. More generally, every measurement basis specifies a pure state ensemble $\mathcal{M} = \{p_s,\ket{\psi_s^{ij}} \}$, where $p_s$ is the probability of outcome $s$ and $\ket{\psi_s^{ij}}$ is the corresponding normalized two-qubit state. The LE then given by,
\begin{equation}
\label{eq:le_gen_def}
 \LE_{ij} = \max_{\mathcal{M}}\sum_s p_sE(\ket{\psi_s^{ij}}), \quad 0\leq\LE_{ij}\leq 1,
\end{equation}
where $E(\ket{\psi_s^{ij}})$ is the entanglement (as measured by the von Neumann entropy of the reduced density matrix) between $i$ and $j$ qubits. Operationally, $\LE_{ij}$ quantifies the maximal entanglement that can be localized between two sites by performing local operations and classical communication (LOCC) on the rest of the system, thus encoding the effective range over which quantum correlations extend \cite{CiracPrl2004,VerstraetePrl2004,PoppPra2005,AmicoRmp2008}. LE between two qubits in a stabilizer state, such as those are appear in this work, is known to be either 0 or 1 \cite{NestPra2004,HeinPra2004,HeinArxiv2006,ZengPra2007}, for examples see \ref{apx:graph_measure}.

Taking inspiration from the classical percolation transition, a natural order parameter for MIPT can be constructed from LE. Let $R$ be the largest separation ratio $|i-j|/(L-1)$ for which the LE remains non-zero,
\begin{equation} 
\orderp = \left\{\text{max}_{i,j} \frac{|i-j|}{L-1}: \; \LE_{ij}>0\right\}, \, 
\label{eq:order_pm}\end{equation}
such that $0 \leq \orderp \leq 1$, as we have assumed open boundary conditions.
In the area-law phase we expect measurements to destroy long-range entanglement, implying finite value of $\text{max}_{i,j}|i-j|$;  hence $\orderp \to 0$ as $L\to \infty$.
In contrast, in the volume-law phase, entanglement persists across the entire system, so $\LE_{ij}$ even at system-scale separations, therefore $\orderp \sim 1$.
\begin{figure}[htbp]
    \centering
    \includegraphics[width=\linewidth]{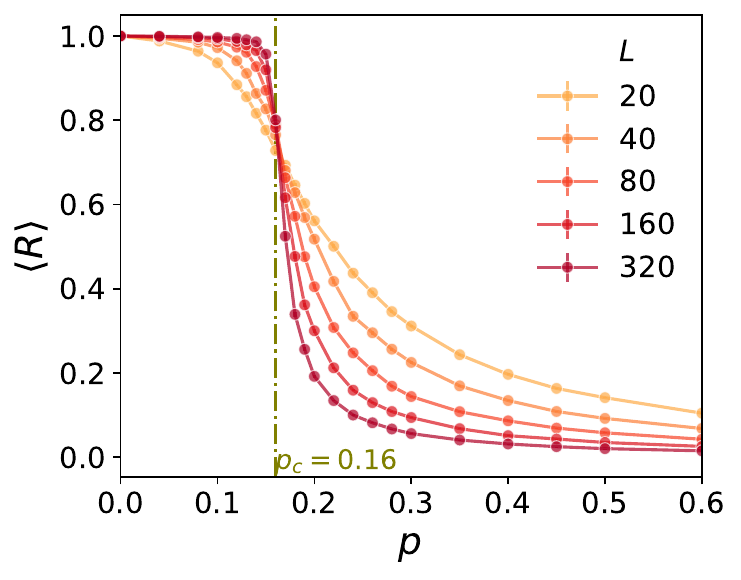}
    \caption{Order parameter {$\avgorderp$} as a function of measurement probability $p$ for random Clifford brickwall circuits of different system sizes $L$. $\avgorderp$ quantifies the maximal normalized separation between qubits with nonvanishing LE, serving as an operational order parameter for the measurement-induced phase transition. The crossing of curves at $p_c \approx 0.16$ signals the transition from the entangling (volume-law) to the disentangling (area-law) phase.}
    \label{fig:LE_p_L}
\end{figure}
Figure~\ref{fig:LE_p_L} illustrates this central result - the order parameter $\avgorderp$ is shown as a function of the measurement probability $p$ for random Clifford brickwall circuits of various system sizes $L$. The $\avgorderp$ is the averaging over different circuit realizations for a fixed probability. The curves for different $L$ cross near $p_c = 0.16$, signaling the measurement-induced transition from the entangling (volume-law) to the disentangling (area-law) phase. This crossing demonstrates that $\avgorderp$ captures the change in the spatial extent of LE across the transition. 
Thus, $\avgorderp$ being strictly zero in one phase and finite in the other directly captures the emergence of a macroscopic long range multipartite entangled cluster  and exhibits non-analytic critical behavior at the transition.  

We argue that $\avgorderp$ serves as a quantum analogue of the percolation order parameter, characterizing the maximal range over which quantum information can be transmitted. \cite{StaufferTaylor1992,GrimmettSpinger2010}.
In classical percolation theory, one studies the emergence of long-range connectivity in a random network as the bond or site occupation probability is increased. In one-dimensional bond percolation, the order parameter \[\theta = \frac{\text{number of nodes in largest cluster}}{\text{total number of nodes}}\] may be interpreted as the length of the largest connected segment normalized by the total system size, thereby quantifying the maximum distance over which classical information can propagate along the chain \cite{StaufferTaylor1992,GrimmettSpinger2010}. Analogously, $\avgorderp$ is strictly zero in the disentangling phase and finite in the entangling phase, reflecting the appearance of long-range quantum connectivity in the form of nonzero LE between distant qubits.

Therefore, $\avgorderp$ also admits a natural operational interpretation in terms of quantum communication: a nonzero value indicates the presence of localizable entanglement between qubits separated by distances scaling with the system size, which in turn guarantees a nonvanishing teleportation fidelity between such distant sites.
This comparison highlights a qualitative distinction between classical and quantum connectivity in one dimension.  In classical one-dimensional bond percolation long-range connectivity is destroyed by any nonzero bond-breaking probability and macroscopic communication is possible only when bonds are never broken corresponding to a critical threshold $p_c = 0$.  In measurement-induced transition, by contrast, long-range quantum connectivity—as quantified by $\avgorderp$—emerges already at a finite measurement rate $p \leq 0.16$. Thus, macroscopic entanglement and long-range quantum communication persist even in the presence of local measurements underscoring the intrinsically quantum nature of the transition.

\begin{figure}[htbp]
    \centering
    \includegraphics[width=\linewidth]{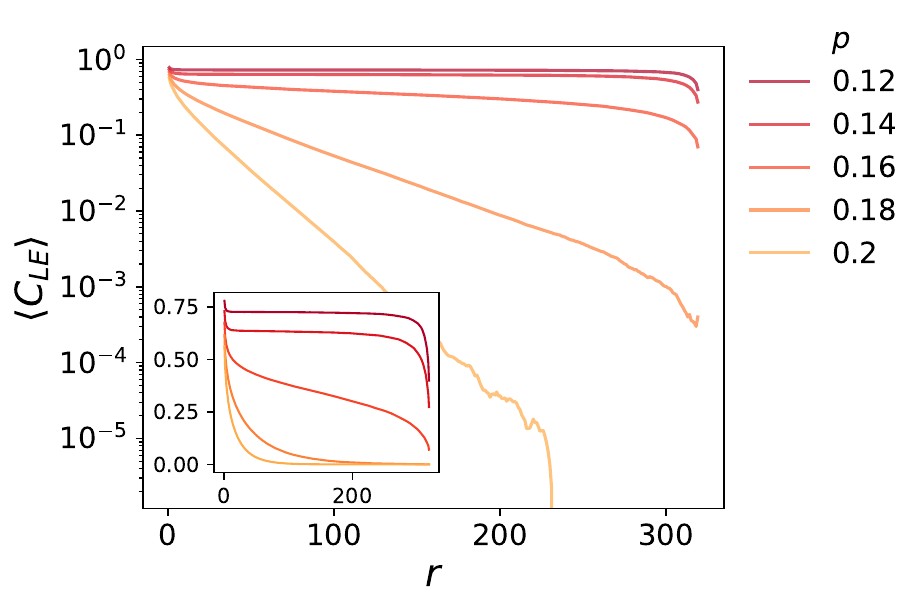}
    \caption{Semilog plot of the entanglement correlation function $\left<\corle(r;p)\right>$. In the area-law phase $(p>p_c)$, the curves are exponential in $r$, enabling extraction of the entanglement correlation length $\xi_E(p)$. At $p<p_c$, the curves approach a constant, reflecting long-range entanglement. The inset shows the same data on linear–linear axes to clearly demonstrate the saturation of $\left<\corle(r;p)\right>$ for $p<p_c$.}
    \label{fig:LE_r_corr_log}
\end{figure}

To characterize entanglement correlations, we compute spatially averaged localizable entanglement (LE) between pairs of qubits separated by a distance $r$. For a fixed circuit realization at measurement probability $p$, this quantity is defined as
\begin{equation}
\label{eq:corr}
\corle(r;p) = \frac{1}{L-r}\sum_{i=1}^{L-r} \LE_{i,i+r},
\end{equation}
where the sum averages over all pairs of sites at separation $r$ along the chain. We then perform an additional average over independent circuit realizations, at a given measurement probability $p$, which we denote by $\left<\corle(r;p)\right>$.

The resulting quantity directly probes the spatial structure of quantum correlations across the system at the measurement-induced transition. As shown in Fig.~\ref{fig:LE_r_corr_log}, in the area-law phase $p>p_c$, the averaged correlation function $\left<\corle(r;p)\right>$ decays exponentially with distance,
\begin{equation}
\left<\corle(r;p)\right> \sim \exp(-r/\xi_E(p)),
    \label{eq:corle_decay}
\end{equation}
thereby defining an entanglement correlation length $\xi_E(p)$ that quantifies the typical spatial range over which quantum correlations persist.

\begin{figure}[htbp]
    \centering
    \includegraphics[width=\linewidth]{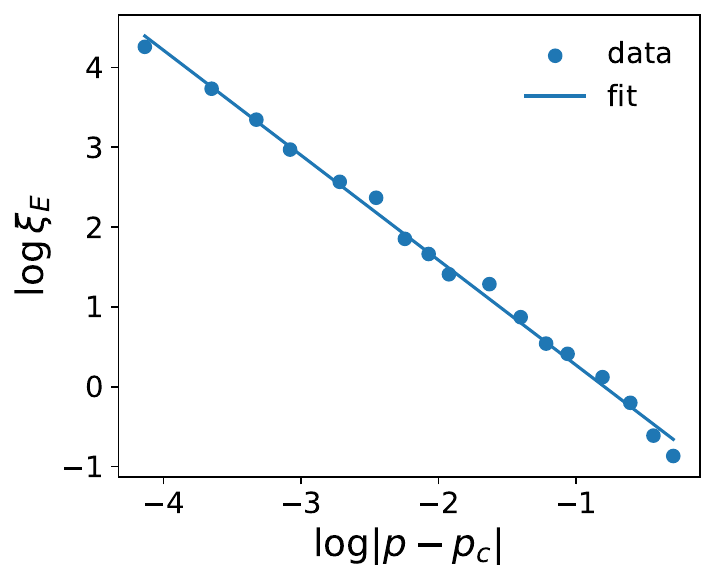}
    \caption{Extraction of the entanglement correlation length in the area-law phase. Log–log plot of the entanglement correlation length $\xi_E(p)$ as a function of the distance from the critical measurement rate, $|p-p_c|$, in the area-law regime $p>p_c$. The data (points) follow a clear power-law divergence, $|p-p_c|^\nu$, and the linear fit (solid line) yields an exponent $\nu=1.31$ consistent with the MIPT.}
    \label{fig:xi_fit_corr}
\end{figure}

The divergence of this length scale upon approaching the transition is captured by a power-law form,
\begin{equation}
\xi_E(p)\sim |p-p_c|^{-\nu},
\end{equation}
with an extracted exponent $\nu = 1.31$ from Fig~\ref{fig:xi_fit_corr}, in excellent agreement with earlier numerical studies on MIPT and consistent with the established universality class\cite{Li_2019,GullansPrl2020}. It is worth mentioning that we have arrived at this exponent from a physically motivated length scale.

In the volume-law phase $(p<p_c)$, $\left<\corle(r;p) \right>$ instead saturates to a finite constant at large separations, reflecting the presence of long-range quantum entanglement.  At the critical measurement rate $p=p_c$, the divergence of $\xi_E(p)$ implies the absence of any intrinsic length scale, and the correlation function becomes scale invariant. At the critical measurement rate $p=p_c$, the divergence of $\xi_E(p)$ signals the emergence of long-range entanglement and the absence of a characteristic microscopic length scale. 
Thus $\left<\corle(r;p)\right>$ provides an independent characterization of MIPT beyond standard entropy-based diagnostics.

We briefly discuss the distinctive features of Clifford circuits and stabilizer states that enable the computation above. For stabilizer states, the LE is restricted to the discrete set $\{0,1 \}$, in contrast to generic quantum states for which LE can vary continuously between 0 and 1. This discreteness reflects the underlying structure of stabilizer states: every stabilizer state is local-Clifford (LC) equivalent to a graph state, and a graph state $\ket{G}$ is specified entirely by a graph $G$ whose vertices label the qubits of $\ket{G}$ \cite{NestPra2004,HeinPra2004,HeinArxiv2006,ZengPra2007}. Under LC operations, graph states transform by local complementations of the underlying graph.  Furthermore, single-qubit Pauli measurements  correspond exactly to vertex deletions (up to known Pauli byproducts) \cite{DahlbergIop2020, DahlbergQuantum2020}.
Therefore, calculating LE between two qubits of a stabilizer state, $|\psi\rangle$, is equivalent to finding whether or not the two vertices in the corresponding graph state $|G\rangle$, are connected by an edge or not after local complementations. For more details and examples, see Appendix~\ref{apx:graph_measure}.

\begin{figure}[htbp]
    \centering
    \includegraphics[width=\linewidth]{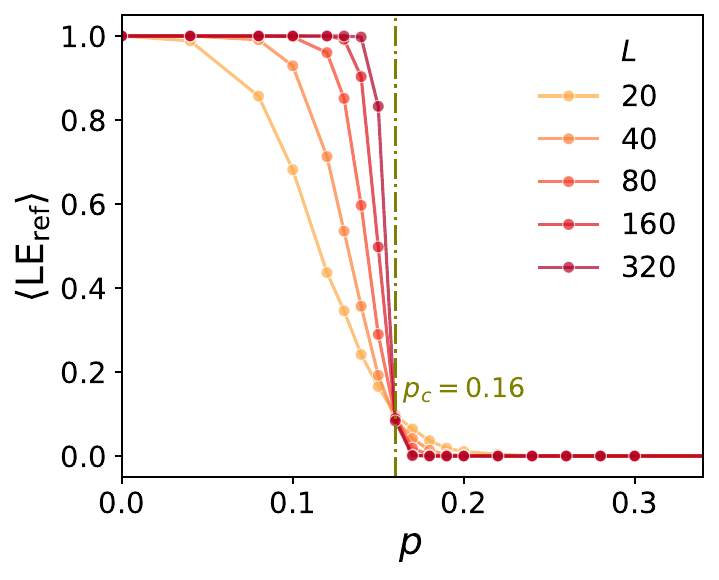}
    \caption{Average LE between a reference qubit and the rest of the system $\left<\leref \right>$, shown as a function of measurement probability $p$ for system sizes  $L=20,40,80,160$. For small measurement rates, $\left<\leref \right> \approx 1$, reflecting the volume-law entangled phase in which the reference qubit shares finite distillable entanglement with the many-body state. As $p$ increases, the curves sharpen with increasing $L$}
    \label{fig:cr1}
\end{figure}
We now discuss application of the above observation to detect MIPT with LE by employing a single reference qubit maximally entangled with the system at the initial time as first proposed in \cite{GullansPrl2020}. 
We find that LE between a reference qubit $R_1$ and the system in Fig~\ref{fig:cr1},  sharply distinguishes the two phases: it is finite for $p<p_c$ and vanishes for $p>p_c$, as guaranteed by the theorem discussed, with the same critical point $p_c$ extracted from the many-body LE.

\begin{figure}[htbp]
    \centering
    \includegraphics[width=\linewidth]{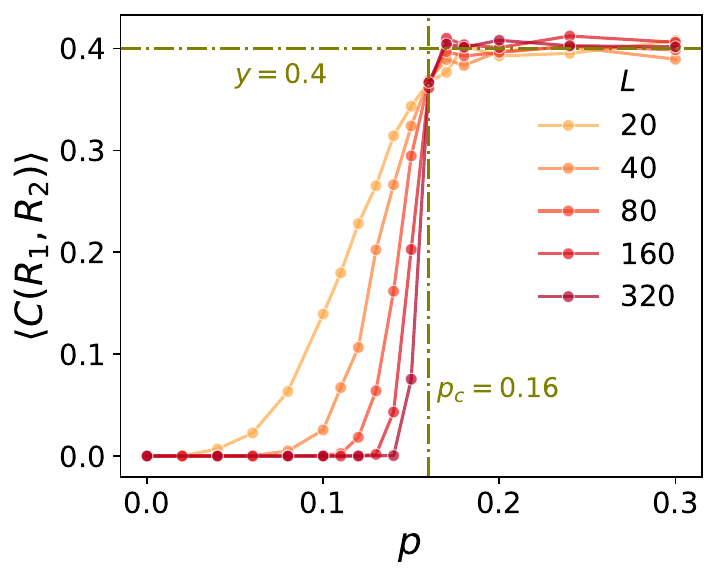}
    \caption{Concurrence of two reference qubits across the measurement-induced transition. Ensemble-averaged concurrence $\avgconr$ as a function of measurement rate $p$ for system sizes $L=20-320$ (top to bottom  indicated in legend). In the volume-law phase $p<p_c$, monogamy prevents the concentration of entanglement onto the two references, yielding $\avgconr\approx 0$. For $p>p_c$, the monitored dynamics purify the $R_1$, and a fixed nonlocal operation between the references converts this purification into a Bell-like state with $\avgconr \approx \mathcal{O}(1)$. All curves exhibit a sharp rise and scale-invariant behavior near the critical point $p=0.16$(vertical line), and saturate close to the maximal stabilizer-state value $\avgconr =  0.4$ (horizontal dashed line), consistent with the two-qubit stabilizer ensemble. The crossing behavior across system sizes demonstrates that the two-ancilla concurrence provides a scalable and experimentally accessible indicator of the MIPT.}
    \label{fig:cr12}
\end{figure}

We can go further and extend this idea by exploiting entanglement monogamy to construct a probe that uses two reference qubits, $R_1$ and $R_2$, with the advantage that it requires no access to the system for detecting MIPT. The first reference $R_1$ is initially entangled with a specific site of the system, which then undergoes monitored evolution. At the final time we apply a fixed nonlocal unitary on $R_1R_2$ and measure entanglement between them.  We use concurrence as a measure of entanglement as two reference can stay in a mixed state. The behavior of the concurrence is strongly phase dependent. In the volume-law phase, the references and the system remain collectively entangled throughout the monitored evolution. Because of entanglement monogamy, this many-body entanglement cannot be concentrated onto the two reference qubits, and the final two-qubit state remains essentially unentangled, giving $C(R_1,R_2) \approx 0$. In contrast, in the area-law phase, measurements continually purify $R_1$, effectively disentangling it from the system. The final nonlocal gate acting on $R_1$ and $R_2$ then converts this purification into a Bell-like correlated state of the references, producing a concurrence of order unity, $C(R_1,R_2) \approx \mathcal{O}(1)$.
For the numerical study, we simulate monitored dynamics using random Clifford circuits. At the final time step, we apply a non-local CNOT gate between the two reference qubits $R_1$ and $R_2$. The behaviour of $C(R_1,R_2)$ is shown in Fig~\ref{fig:cr12}. purified and thus lies in one of the six single-qubit stabilizer states $\{\ket{0},\,\ket{1},\,\ket{+},\,\ket{-},\,\ket{+i},\,\ket{-i}\}$ prior to the application of the CNOT. Averaging over all possible input states, the resulting concurrence generated by the final non-local gate is
\begin{equation}
    \label{eq:cr12_area}
\left<\conr\right> = \frac{24}{60} = 0.4.
\end{equation}

Thus, the concurrence of two isolated ancillas provides a efficient, fully operational diagnostic of MIPT. Furthermore, the concurrence of any two-qubit state can be reconstructed solely from the Pauli correlators $G_{\alpha\beta} = \expt{\sigma_{1\alpha}\sigma_{2\beta}} = \Tr \left(\sigma_{1\alpha}\sigma_{2\beta}\rho\right)$  where $\alpha,\beta \in \{x,y,z\}$ \cite{WangPra2002}. Consequently, our protocol requires measurements only on the two reference qubits and treats the many-body system entirely as a black box, making it especially suitable for near-term experimental platforms. In fact, determining whether $R_1$ and $R_2$  become entangled or remain separable is already sufficient to distinguish the two dynamical phases and thereby probe the MIPT.

In summary, we have discovered an order parameter to characterize MIPT based on LE that has an operational interpretation in terms of   quantum teleportation between distant sites in close analogy with how classical information propagates in classical networks.  Thus, unlike classical one-dimensional percolation where macroscopic connectivity exists only at zero bond breaking, the measurement-induced transition exhibits finite-rate emergence of long-range quantum connectivity—signaled by a nonzero $\avgorderp$—enabling macroscopic localizable entanglement and nonvanishing teleportation fidelity despite local measurements.
By exploiting the stabilizer–graph-state correspondence, we directly demonstrate an emergent entanglement correlation length that diverges at the critical measurement rate. These results provide a clear operational interpretation of quantum-information transport in monitored circuits, complementing and extending entropy-based diagnostics.

Building on this framework, we introduced a monogamy-based two-ancilla probe that detects the transition solely through two-qubit measurements, without requiring access to the many-body system. The concurrence between the ancillas exhibits a sharp rise across the transition, switching from essentially zero in the volume-law phase to a finite value in the area-law phase.

Together, these advances provide a unified conceptual picture—linking teleportation-based interpretations of monitored circuits, the classical percolation viewpoint of MIPTs, and possibly more experimentally accessible probes on current quantum platforms. The operational nature of our tools opens the door to systematic exploration of quantum networks and enables direct percolation-inspired mappings within quantum circuits, allowing universality to be tested across a broad class of hybrid quantum systems.

\begin{acknowledgments}
The authors would like to thank HPCE, IIT Madras, for providing the computational facility for numerical simulations. 
  This work was supported in part by grant  DST/ICPS/QusT/Theme-3/2019/Q69 and New faculty Seed Grant from IIT Madras. The authors were supported, in part, by a grant from Mphasis to the Centre for Quantum Information, Communication, and Computing (CQuICC) at IIT Madras.
\end{acknowledgments}

\bibliography{apssamp}


\appendix

\section{Graph Transformations Under Pauli Measurements}
\label{apx:graph_measure}
Graph states provide a convenient graphical way to describe multi-qubit 
entanglement.  Each qubit corresponds to a vertex of a graph $G=(V,E)$, and 
a controlled-$Z$ (CZ) gate is applied on every edge.  
Once the graph state is prepared, the effect of measuring any qubit in the 
$Z$, $X$, or $Y$ basis can be understood directly in terms of simple 
modifications of the underlying graph.  
These modifications are known as \emph{local complementations}.  
Below we summarize the measurement rules in the simplest possible way.

\subsection{$Z$-measurement: remove the qubit}

If a qubit $v$ is measured in the $Z$ basis, it becomes disentangled from the 
rest of the system.  Graphically:

\begin{quote}
\textbf{Rule 1.} 
A $Z$-measurement on $v$ deletes the vertex $v$ and all edges incident to it.
\end{quote}

Thus the remaining qubits form the subgraph $G-v$.

\subsection{Local complementation: the basic rewiring rule}

Measurements in the $X$ or $Y$ basis are equivalent to performing a 
basis rotation on the measured qubit followed by a projective $Z$ 
measurement.  
The basis rotation affects only the connections \emph{between the neighbors} 
of the measured qubit.  
This is captured by the graph operation called \emph{local complementation}.

\begin{quote}
\textbf{Local complementation $\tau_v(G)$:}  
Look at all neighbors of $v$.  
For every pair of neighbors, toggle the edge between them:\\
— If an edge is present, remove it;\\
— If no edge is present, add it.\\
All other edges remain unchanged.
\end{quote}

Local complementation rewires only the neighborhood of $v$; it does not 
affect the rest of the graph.

\subsection{$X$-measurement rule}

An $X$-basis measurement on vertex $v$ corresponds to one local 
complementation followed by deletion:

\begin{quote}
\textbf{Rule 2 (X-measurement).}
\[
M_X(v): \qquad G' = \tau_v(G) - v .
\]
\end{quote}

Thus the neighbors of $v$ are first rewired according to $\tau_v$, and then 
$v$ is removed from the graph.

\subsection{$Y$-measurement rule}

A $Y$-measurement involves an additional phase rotation.  
This produces one extra local complementation on any neighbor of $v$.

\begin{quote}
\textbf{Rule 3 (Y-measurement).}
\[
M_Y(v): \qquad 
G' = \tau_u\!\left(\tau_v(G)\right) - v,
\qquad u\in N(v).
\]
Any choice of $u$ yields a graph that is locally Clifford–equivalent 
(i.e., physically equivalent up to single-qubit basis changes).
\end{quote}

Thus a $Y$ measurement consists of two local complementations followed by 
vertex deletion.

\subsection{Summary}

\begin{itemize}
    \item $Z$-measurement: remove the vertex.
    \item $X$-measurement: rewiring via $\tau_v$, then remove the vertex.
    \item $Y$-measurement: rewiring via $\tau_v$ and a second $\tau_u$ 
          (with $u\in N(v)$), then remove the vertex.
\end{itemize}

These simple rules allow one to track the post-measurement entanglement 
structure directly from the graph, without manipulating operators or state 
vectors.  They are particularly useful in measurement-based quantum 
computation, stabilizer circuits, and measurement-induced phase transitions.

\subsection{Examples with Graphical Representations}

Below we illustrate the three Pauli measurement rules using simple graph
examples. 

\subsubsection*{Example 1: $Z$-measurement on a chain}

Consider the 3-qubit chain $1{-}2{-}3$.

\begin{center}
\begin{tikzpicture}[scale=1.0, every node/.style={circle,draw,inner sep=2pt}]
\node (1) at (0,0) {1};
\node (2) at (1.5,0) {2};
\node (3) at (3,0) {3};
\draw (1)--(2)--(3);
\end{tikzpicture}
\end{center}

Measuring qubit $2$ in the $Z$ basis simply removes vertex $2$:

\[
M_Z(2): \quad G' = G - 2.
\]

\begin{center}
\begin{tikzpicture}[scale=1.0, every node/.style={circle,draw,inner sep=2pt}]
\node (1) at (0,0) {1};
\node (3) at (1.5,0) {3};
\end{tikzpicture}
\end{center}

Thus the two endpoints become disconnected.

\bigskip


\subsubsection*{Example 2: $X$-measurement on a chain graph and  star graph}
Consider the 3-qubit chain $1{-}2{-}3$.

\begin{center}
\begin{tikzpicture}[scale=1.0, every node/.style={circle,draw,inner sep=2pt}]
\node (1) at (0,0) {1};
\node (2) at (1.5,0) {2};
\node (3) at (3,0) {3};
\draw (1)--(2)--(3);
\end{tikzpicture}
\end{center}

Measuring, qubit 2 in the $X$.

Step 1: apply local complementation $\tau_2$, which toggles the edge between neighbous $\{1,3\}$.

\begin{center}
\begin{tikzpicture}[scale=1.0, every node/.style={circle,draw,inner sep=2pt}]
\node (1) at (0,0) {1};
\node (2) at (1.5,0) {2};
\node (3) at (3,0) {3};
\draw (1)--(2)--(3);
\draw (1) to[out=45,in=135] (3);
\end{tikzpicture}
\end{center}

Step 2: Remove qubit 2

\begin{center}
\begin{tikzpicture}[scale=1.0, every node/.style={circle,draw,inner sep=2pt}]
\node (1) at (0,0) {1};
\node (3) at (3,0) {3};
\draw (1)--(3);
\end{tikzpicture}
\end{center}

Consider the star graph with center $4$:

\begin{center}
\begin{tikzpicture}[scale=1.0, every node/.style={circle,draw,inner sep=2pt}]
\node (4) at (0,0) {4};
\node (1) at (-1.2,1) {1};
\node (2) at (-1.2,-1) {2};
\node (3) at (1.2,0) {3};
\draw (4)--(1);
\draw (4)--(2);
\draw (4)--(3);
\end{tikzpicture}
\end{center}

Step 1: apply local complementation $\tau_4$, which toggles all edges among $\{1,2,3\}$, producing a triangle:

\begin{center}
\begin{tikzpicture}[scale=1.0, every node/.style={circle,draw,inner sep=2pt}]
\node (4) at (0,0) {4};
\node (1) at (-1.2,1) {1};
\node (2) at (-1.2,-1) {2};
\node (3) at (1.2,0) {3};
\draw (4)--(1);
\draw (4)--(2);
\draw (4)--(3);
\draw (1)--(2)--(3)--(1);
\end{tikzpicture}
\end{center}

Step 2: delete the measured vertex:

\[
M_X(4): \quad G' = \tau_4(G) - 4.
\]

\begin{center}
\begin{tikzpicture}[scale=1.0, every node/.style={circle,draw,inner sep=2pt}]
\node (1) at (-1.2,1) {1};
\node (2) at (-1.2,-1) {2};
\node (3) at (1.2,0) {3};
\draw (1)--(2)--(3)--(1);
\end{tikzpicture}
\end{center}

The three remaining qubits form a complete graph.

\bigskip


\subsubsection*{Example 3: $Y$-measurement on a chain}

Again consider the chain $1{-}2{-}3$.  
We measure qubit $2$ in the $Y$ basis.

\begin{center}
\begin{tikzpicture}[scale=1.0, every node/.style={circle,draw,inner sep=2pt}]
\node (1) at (0,0) {1};
\node (2) at (1.5,0) {2};
\node (3) at (3,0) {3};
\draw (1)--(2)--(3);
\end{tikzpicture}
\end{center}

Step 1: apply $\tau_2$, which toggles the edge between neighbors $\{1,3\}$:

\begin{center}
\begin{tikzpicture}[scale=1.0, every node/.style={circle,draw,inner sep=2pt}]
\node (1) at (0,0) {1};
\node (2) at (1.5,0) {2};
\node (3) at (3,0) {3};
\draw (1)--(2)--(3);
\draw (1) to[out=45,in=135] (3);
\end{tikzpicture}
\end{center}

Step 2: apply $\tau_1$ (we may choose any neighbor).  
This removes the edge $(2,3)$ but keeps $(1,3)$.

Step 3: delete vertex $2$:

\[
M_Y(2) :\quad G' = \tau_1(\tau_2(G)) - 2.
\]

\begin{center}
\begin{tikzpicture}[scale=1.0, every node/.style={circle,draw,inner sep=2pt}]
\node (1) at (0,0) {1};
\node (3) at (1.5,0) {3};
\draw (1)--(3);
\end{tikzpicture}
\end{center}

Thus a $Y$ measurement on the middle qubit \emph{connects the endpoints}.

\bigskip

\subsubsection*{Example 4: $Y$-measurement on a 4-cycle}

Consider the 4-qubit cycle (square)
\[
1 - 2 - 3 - 4 - 1 .
\]

\begin{center}
\begin{tikzpicture}[scale=1.1, every node/.style={circle,draw,inner sep=2pt}]
\node (1) at (0,1) {1};
\node (2) at (1.5,1) {2};
\node (3) at (1.5,-0.3) {3};
\node (4) at (0,-0.3) {4};
\draw (1)--(2)--(3)--(4)--(1);
\end{tikzpicture}
\end{center}

We measure qubit $1$ in the $Y$ basis.  
This involves two local complementations followed by deletion.

\paragraph*{Step 1: Apply $\tau_1$.}  
The neighbors of $1$ are $\{2,4\}$.  
Local complementation toggles the edge $(2,4)$, adding it:

\begin{center}
\begin{tikzpicture}[scale=1.1, every node/.style={circle,draw,inner sep=2pt}]
\node (1) at (0,1) {1};
\node (2) at (1.5,1) {2};
\node (3) at (1.5,-0.3) {3};
\node (4) at (0,-0.3) {4};
\draw (1)--(2)--(3)--(4)--(1);
\draw (2)--(4);   
\end{tikzpicture}
\end{center}

\paragraph*{Step 2: Apply $\tau_2$.}  
We may choose either neighbor $2$ or $4$; both give LC-equivalent results.  
Choosing $u=2$, its neighbors are now $\{1,3,4\}$.  
Local complementation toggles edges among $\{1,3,4\}$:
\[
(1,3): 0\to 1, \qquad
(1,4): 1\to 0, \qquad
(3,4): 1\to 0 .
\]

The resulting graph is:

\begin{center}
\begin{tikzpicture}[scale=1.1, every node/.style={circle,draw,inner sep=2pt}]
\node (1) at (0,1) {1};
\node (2) at (1.5,1) {2};
\node (3) at (1.5,-0.3) {3};
\node (4) at (0,-0.3) {4};
\draw (1)--(2);
\draw (2)--(3);
\draw (2)--(4);
\draw (1)--(3); 
\end{tikzpicture}
\end{center}

\paragraph*{Step 3: Delete vertex 1.}

Removing vertex $1$ and all edges incident to it gives:

\[
M_Y(1): \quad G' = \tau_2(\tau_1(G)) - 1.
\]

\begin{center}
\begin{tikzpicture}[scale=1.1, every node/.style={circle,draw,inner sep=2pt}]
\node (2) at (0,1) {2};
\node (3) at (1.5,1) {3};
\node (4) at (0.75,-0.3) {4};
\draw (2)--(3);
\draw (2)--(4);
\end{tikzpicture}
\end{center}

The final graph is a ``V'' shape $2{-}4$ and $2{-}3$, with qubit $1$ removed.
This illustrates how a $Y$ measurement on a vertex of a cycle can break the
cycle while creating new correlations among its neighbors.

\end{document}